\documentclass[12pt,aps,roupedaddress,aa,showkeys]{revtex4}

\usepackage{color}
\usepackage{latexsym}

\usepackage{amsmath}
\usepackage{graphicx}

\usepackage{dcolumn}

\input epsf

\def\lesssim{\mathrel{\hbox{\rlap{\hbox
{\lower4pt\hbox{$\sim$}}}\hbox{$<$}}}}
 
\def\gtrsim{\mathrel{\hbox{\rlap{\hbox
{\lower4pt\hbox{$\sim$}}}\hbox{$>$}}}}

\begin{document}

\title{Solving 1-D Special Relativistic Hydrodynamics(SRH) Equations Using 
Different Numerical Method and Results from Different Test Problems}
	
\author
{Orhan D\"{o}nmez\footnote{electronic address:odonmez@nigde.edu.tr}}

\address{Nigde University Faculty of Art and Science, 
Physics Department, Nigde, Turkey} 


\date{\today}

\begin{abstract}
In this paper, we have solved $1D$ special relativistic hydrodynamical equations 
using different numerical method in computational gas dynamics. The numerical solutions
of these equations for smooth wave cases give better solution when we use 
$Non-TVD$(Total Variable Diminishing) but solution of discontinuity wave produces some
oscillation behind the shock. On the other hand, $TVD$ type schemes give good 
approximation at discontinuity cases. Because $TVD$ schemes completely remove  
the oscillations, they reduce locally the accuracy of the solution around
the extrema. 

\end{abstract}


\maketitle

\section{INTRODUCTION}

The invention of the digital computer and its introduction into the world of science
and technology has led to the development, and increased awareness, of the concept of
approximation. This concerns the theory of numerical approximation of a set of 
equations, taken as a mathematical model of a physical system. However, it also 
concerns the notion of approximation involved in the definition of this mathematical 
model with respect to the complexity of physical world.We are concerned here with 
physical systems for which is assumed that the basic equations describing their 
behavior is known theoretically  but for which no analytic solutions exist, and 
consequently an approximate numerical solution will be sought instead.The 
approximation is relative to a given time and environment, and these are being 
extended with the evolution of computer technology.We can state that a mathematical 
model for the behavior of a astrophysical system, and in particular the system of 
fluid flows, can only be defined after consideration of the level of the 
approximation required in order to achieve an acceptable accuracy on a defined set 
of dependent and independent variables.For instance, evolution of relativistic 
hydrodynamical system can be considered to depend on conserve and primitive 
variables. 

Actually, physicists propose various levels of description of our physical world, 
ranging from subatomic or molecular, microscopic or macroscopic up to the 
astronomical scale. So fluid dynamics is essentially the study of the interactive 
motion and behavior  of a large number of individual elements. From this point of 
view we understand easily why concept of fluid mechanics can be applied large 
number of interacting elements, such as astrophysical phenomenon.

The astrophysical problems creates strong shocks region due to strong gravitational 
field. An accurate description of relativistic cases with strong shocks is needed 
for study of important problems, such as accreting of compact objects, stellar 
collapse, and coalescing of compact binaries. At this end, we have started testing 
different numerical methods to solve the relativistic hydrodynamical equations.

In this paper, first we introduce the special relativistic hydrodynamical(SRH) 
equation and their components. Second, we give detail discussion about numerical 
schemes we have used here. Finally, we discuss numerical solution of SRH equation from 
different numerical schemes when we applied them to the 
different test problems.

\section{FORMULATION}
\label{formulation}

The General Relativistic Hydrodynamic (GRH) equations in  Refs.
\cite{Don_Fon} and \cite{Fon_Mil}, written in the standard
covariant form, consist  of the local conservation laws of the
stress-energy tensor $T^{\mu \nu }$  and the matter current density $
J^\mu$:

\begin{eqnarray}
\bigtriangledown_\mu T^{\mu \nu} = 0 ,\;\;\;\;\; 
\bigtriangledown_\mu J^\mu = 0.
\label{covariant derivative}
\end{eqnarray}

\noindent
Greek indices run from $0$ to $3$, Latin indices from $1$ to $3$, and units 
in which the speed of light $c = 1$ are used.

Defining the characteristic waves of the general
relativistic hydrodynamical equations is not trivial with imperfect
fluid stress-energy tensor. We neglect the viscosity and heat
conduction effects.  This defines the  perfect fluid
stress-energy tensor. We use this stress-energy tensor to derive the
hydrodynamical equations. With this 
perfect fluid stress-energy tensor, we can solve some problems which
are  solved by the Newtonian hydrodynamics with viscosity, such as
those involving angular momentum transport and shock waves on an
accretion disk, etc. Entropy for 
perfect fluid is conserved along the fluid lines. The stress energy tensor
for a perfect fluid is given as

\begin{equation}
T^{\mu \nu} = \rho h u^\mu u^\nu + P g^{\mu \nu}.
\label{des 7}
\end{equation}

\noindent
A perfect fluid  is a fluid that moves through spacetime with a
4-velocity $u^{\mu}$ which may vary from event to event. It exhibits a
density of mass $\rho$ and isotropic pressure $P$ in the rest frame of
each fluid element. $h$ is the specific
enthalpy, defined as

\begin{equation} 
h = 1 + \epsilon +\frac{P}{\rho}.
\label{hdot}
\end{equation}

\noindent
Here  $\epsilon$ is the specific internal energy. The equation of
state might have  the 
functional form $P = P(\rho, \epsilon)$. The perfect
gas equation of state, 

\begin{equation}
P = (\Gamma -1 ) \rho \epsilon,
\label{flux split21}
\end{equation}

\noindent
is such a functional form.

The conservation laws in the form given in Eq.(\ref{covariant
derivative}) are not suitable for 
the use in advanced numerical schemes. In order to carry out numerical
hydrodynamic evolutions such as those reported in \cite{Fon_Mil}, and to
use high resolution shock capturing schemes, the hydrodynamic equations 
after the 3+1 split
must be written as a hyperbolic system of first order flux
conservative equations. We write Eq.(\ref{covariant derivative}) in
terms of coordinate derivatives, using the coordinates ($x^0 = t, x^1,
x^2, x^3$). Eq.(\ref{covariant derivative}) is projected onto the
basis $\lbrace n^\mu, (\frac{\partial}{\partial x^i})^\mu \rbrace$,
where $n^\mu$ is a unit timelike vector normal to a given
hypersurface. After a straightforward calculation and neglecting the GR 
part of equation
we get in $1D$ (see ref.\cite{Fon_Mil}),

\begin{equation}
\partial_t \vec{U} + \partial_x \vec{F}^x = 0, 
\label{desired equation}
\end{equation} 

\noindent
where $\partial_t = \partial / \partial t$ and $\partial_x = \partial
/ \partial x$. This basic step serves to identify the
set of unknowns, the vector of conserved quantities $\vec{U}$, and
their corresponding fluxes $\vec{F}^x(\vec{U})$. With the
equations in conservation form, almost every high
resolution method devised to solve hyperbolic systems of conservation
laws can be extended to GRH.

The evolved state vector $\vec{U}$ consists of  the conservative
variables $(D, S_x, \tau)$ which are conserved variables for density,
momentum and energy respectively; in terms of the
primitive variables $(\rho, v^x, \epsilon)$, this becomes \cite{Fon_Mil}

\begin{equation}
\vec{U} = \left( 
\begin{array}{c} 
D \\
S_x \\
\tau 
\end{array} 
\right) = \left( 
\begin{array}{c}
\sqrt{\gamma} W \rho \\
\sqrt{\gamma}\rho h W^2 v_x \\
\sqrt{\gamma} (\rho h W^2 - P - W \rho) 
\end{array} 
\right). 
\label{matrix form of conserved quantities}
\end{equation}

\noindent
Here $\gamma$ is the determinant of the 3-metric $\gamma_{xj}$ which is a unit 
matrix for special relativity, $v_x$ is the fluid 3-velocity in $x$ direction, 
and W is the Lorentz factor, 

\begin{equation}
W = \alpha u^0 = (1 - \gamma_{xj} v^x v^j)^{-1/2}. 
\label{Wdot1}
\end{equation}

\noindent 
The flux vectors $\vec{F^x}$ are given by \cite{Fon_Mil}

\begin{equation}
\vec{F}^x =  \left( \begin{array}{c}
\alpha v^x D \\
\alpha \lbrace v^x S_j + \sqrt{\gamma} P
\delta ^{x}_{j} \rbrace    \\
\alpha \lbrace v^x \tau + \sqrt{\gamma}
v^x P \rbrace \end{array} \right). 
\label{matrix form of Flux vector}
\end{equation}

\noindent
The spatial components of the 4-velocity $u^x$ are related to the
3-velocity by the following formula: $u^x = W v^x $. $\alpha$, which equals $1$ for
special relativistic case, is the lapse function of the spacetime.

The use of HRSC  scheme requires the
spectral decomposition of the Jacobian matrix of the
system, $\partial \vec{F^x} / \partial \vec{U}$. 
The spectral decomposition of the Jacobian matrices of the SRH 
 equations with a general equation of state
was reported in \cite{Fon_Mil}.  

We started the solution by considering an equation of state in which the
pressure $P$ is a function of $\rho$ and $\epsilon$, $P = P(\rho,
\epsilon)$. The relativistic speed of sound in the fluid $C_s$ is
given by \cite{Fon_Mil}

\begin{equation}
C_{s}^2 = \left. \frac{\partial P}{\partial E} \right|_S. =  \frac{\chi}{h} + \frac{P
\kappa}{\rho^2 h},   
\label{cdot}
\end{equation}

\noindent
where $\chi = \partial P / \partial \rho\vert_\epsilon$, $\kappa = \partial
P / \partial \epsilon\vert_\rho$, $S$ is the entropy per particle, and
$E = \rho + \rho \epsilon$ is the total rest energy density. 

In order to use numerical schemes to solve SRH equation, eigenvalues and left and 
right eigenvector must be defined for the Jacobian matrix. A complete set of 
the right and left eigenvectors $[\vec{r}_i]$ and  corresponding
eigenvalues $\lambda_i$  along the $x$-direction is given in \cite{DONMEZ1}.

In any relativistic hydrodynamics code evolving the conserved
quantities $(D, S, \tau)$ in time, the primitive variables $(P,
\rho,v)$ have to be computed  from the conserved quantities at least
once per time step. In our code, this is achieved using relations
(\ref{flux split21}),(\ref{hdot}),(\ref{matrix form of conserved
quantities}) and (\ref{Wdot1}) to construct the function \cite{MM}

\begin{equation}
f(P) = (\Gamma -1) \rho_* \epsilon_* - P,
\label{funcdot}
\end{equation}

\noindent
where  Eq. (\ref{matrix form of conserved quantities}) gives 

\begin{equation}
\rho_* = \frac{D}{\sqrt{\gamma} W_*}
\label{dens}
\end{equation}

\noindent
and Eqs. (\ref{hdot}) and (\ref{matrix form of conserved
quantities}) give 

\begin{equation}
\epsilon_* = \frac{\tau + D (1 - W_*) + \sqrt{\gamma} (1 - W_*{^2})
P}{D W_*}. 
\label{energy}
\end{equation}

\noindent
Here
\begin{equation}
W_* = \frac{1}{\sqrt{1 - v^2}},
\label{lore}
\end{equation}

\noindent
and $v^2 = \gamma^{jk} v_j v_k \;= v_j v^j$. 

\noindent
From Eq.(\ref{matrix form of conserved quantities}), the
following relation between $P$, $v$,  and the conserved quantities can be derived: 

\begin{equation}
v_j = \frac{S_j}{\tau + \sqrt{\gamma} P + D}.
\label{vdot}
\end{equation}

\noindent
From Eqs.(\ref{lore}) and (\ref{vdot}), we get

\begin{equation}
W_* = \frac{1}{\sqrt{1 - \frac{S_j}{\tau + \sqrt{\gamma} P + D}
\gamma^{jk}\frac{S_k}{\tau + \sqrt{\gamma} P + D}}}.   
\label{lorentz}
\end{equation}

Setting $f(P) = 0$ in equation (\ref{funcdot}) gives a nonlinear
implicit equation for $P$.  It can be solved
using a root finding method; in this work, we are using the
false-position method \cite{NUMREC}. The zero of $f(P)$ in the
physically allowed domain $P_{min} < P < P_{max}$ 
determines the pressure, and the monotonicity of $f(P)$ in that domain
ensures the uniqueness of the solution \cite{MM}. The lower bound of the
physically allowed domain $P_{min}$, defined by $P_{min} = |S - \tau
-D|$, is obtained from (\ref{vdot}) by taking into account that (in our 
units) $|v| \leq 1$, and $P_{max}$ can be taken to have any
sufficiently large value. Knowing $P$, Eq.(\ref{vdot}) then directly
gives $v$, while the remaining state quantities are  obtained in a
straightforward manner from Eqs.(\ref{hdot}), (\ref{matrix form of
conserved quantities}), and (\ref{Wdot1}).


\section{Numerical Schemes and Method}
\label{Numerical Schems}

The special relativistic hydrodynamical equations in $1D$  can be
written in the form

\begin{equation}
\frac{\partial \vec{U}}{\partial t} + \frac{\partial \vec{F}^x}{\partial
x}  = 0,
\label{conserved differantial equation}
\end{equation}  

\noindent
Discretization of the hydrodynamical equations (\ref{conserved differantial
equation}) gives

\begin{eqnarray}
\frac{\partial \vec{U}_{i}}{\partial t} + \frac{(\vec{f^*})_{i+1/2} - 
(\vec{f^*})_{i-1/2}}{\bigtriangleup x}  = 0.
\label{discreization equation}
\end{eqnarray} 

\noindent
where $(\vec{f^*})_{i+1/2}$ is the numerical flux calculated  at
the interfaces $i\pm 1/2$ of spatial cell $i$.

In here, we will explain the numerical methods we use to solve
the hydrodynamical equations. First, we will introduce the flux splitting
method in which  fluxes are defined depending on the sign of
eigenvalues of Jacobian matrix which is defined from SRH
equations. Second, we will explain the MUSCL-type schemes, in which
the state  
variables at the interfaces are obtained from an extrapolation between
neighboring  cell averages.


\subsection{Flux Split Method}
\label{Flux Split Method}

First, we consider the flux splitting method, in which the flux is 
decomposed into the  
part contributing to the eigenfields with positive eigenvalues (fields
moving to the right) and the part with negative eigenvalues (fields
moving to the left) \cite{HIRSCH,Toro}. These fluxes are then
discretized with one-sided or upwind differences depending on the sign
of the particular eigenvalue. For example, the flux of material moving
in the $+x$ direction is differenced with a backward spatial difference.

For the flux split method, one assumes that \cite{HIRSCH,Toro}

\begin{equation}
\vec{F}^x(\zeta \vec{U}) = \zeta \vec{F}^x(\vec{U}),
\label{flux split1}
\end{equation}

\noindent
for any constant $\zeta$. This only holds for the fluxes of
Eq.(\ref{matrix form of Flux vector}) if the equation of state has the
functional form $P = P(\rho, \epsilon) = \rho f(\epsilon)$, for some
function $f(\epsilon)$. Therefore, we use the perfect gas equation of
state, 

\begin{equation}
P = (\Gamma -1 ) \rho \epsilon,
\label{flux split2}
\end{equation}

\noindent
where $\Gamma$ is the adiabatic index of the fluid. From the
Eq.(\ref{flux split1}) , the flux vector $\vec{F}^x$ can be written 

\begin{equation}
\vec{F}^x = (\frac{\partial \vec{F}^x}{\partial \vec{U}}) \vec{U}.
\label{flux split3}
\end{equation}

\noindent
Using the spectral decomposition, one can write the Jacobian matrix
$\partial \vec{F}^x/\partial \vec{U}$ in the form \cite{HIRSCH,Toro} 

\begin{equation}
\frac{\partial \vec{F}^x}{\partial \vec{U}}  = (M^x) \Lambda^x (M^x)^{-1},
\label{flux split4}
\end{equation}

\noindent
where $M^x$ is the matrix whose columns are the right eigenvectors of
the system in the x-direction, and $\Lambda^x$ is a diagonal matrix
constructed from the corresponding eigenvalues which are given in \cite{DONMEZ1}. 

Next, we split the flux into the part that is moving to the right and
the part that is moving to the left. Using Eqs.(\ref{flux split3}) and
(\ref{flux split4}) this gives \cite{HIRSCH,Toro} 

\begin{eqnarray}
\vec{F}^x = (\vec{F}^x)^+ + (\vec{F}^x)^- = \{(M^x) (\Lambda^x)^+
(M^x)^{-1}\}\vec{U} + \nonumber \\
\{(M^x) (\Lambda^x)^- (M^x)^{-1}\}\vec{U},
\label{flux split5}
\end{eqnarray}

\noindent
where $(\Lambda^x)^+ = \frac{1}{2} (\Lambda^x + |\Lambda^x|)$, and
$(\Lambda^x)^- = \frac{1}{2} (\Lambda^x - |\Lambda^x|)$. If we use a
first-order upwind flux, we define   

\begin{eqnarray}
(\vec{f}^*_{i+1/2}) = (\vec{F}^x)^{+}_{i} + (\vec{F}^x)^{-}_{i+1},
\label{flux split6}
\end{eqnarray}

\noindent
and

\begin{eqnarray}
(\vec{f}^*_{i-1/2}) = (\vec{F}^x)^{+}_{i-1} + (\vec{F}^x)^{-}_{i}.
\label{flux split7}
\end{eqnarray}

\noindent
When these are substituted into Eq.(\ref{discreization equation}), we
get

\begin{equation}
\vec{U}^{n+1}_{i} = \vec {U}^{n}_{i} - \frac{ \bigtriangleup
t}{\bigtriangleup x} [(\vec{F}^x)^{+}_{i} + (\vec{F}^x)^{-}_{i+1} -
((\vec{F}^x)^{+}_{i-1} + (\vec{F}^x)^{-}_{i})]^n . 
\label{flux split8}
\end{equation}

\noindent
This scheme is first-order accurate in space and time.

Second order accurate  flux-splitting method can also
be constructed; see \cite{DFIM}.

\subsection{MUSCL-Type Methods }
\label{MUSCL-Type Methods }

We introduce HRSC(High Resolution Shock
Capturing) schemes which use slope limiters to kill spurious
oscillations, called MUSCL-type schemes.  
MUSCL stands for Monotone Upstream-centered Scheme for Conservation
Laws. The MUSCL-type scheme allows us to construct higher order
methods, fully discrete, semi-discrete and also implicit
methods \cite{HIRSCH,Toro}. While higher order linear schemes produce
spurious oscillations, 
the MUSCL-type scheme achieves a high order of accuracy by data
reconstruction, where the reconstruction is constrained so as to avoid
spurious oscillations.

The value of any quantity, $u^{n}_{i}$
represents an integral average in cell
$[x_{i-\frac{1}{2}},x_{i+\frac{1}{2}}]$, given by

\begin{eqnarray}
u^{n}_{i} = \frac{1}{\Delta x}
\int_{x_{i-\frac{1}{2}}}^{x_{i+\frac{1}{2}}} u(x,t^n) dx.
\label{godunov integral}
\end{eqnarray}

\begin{center}
\begin{figure}[ht]
\epsfxsize=15.cm \epsfysize=15.cm
\epsfbox{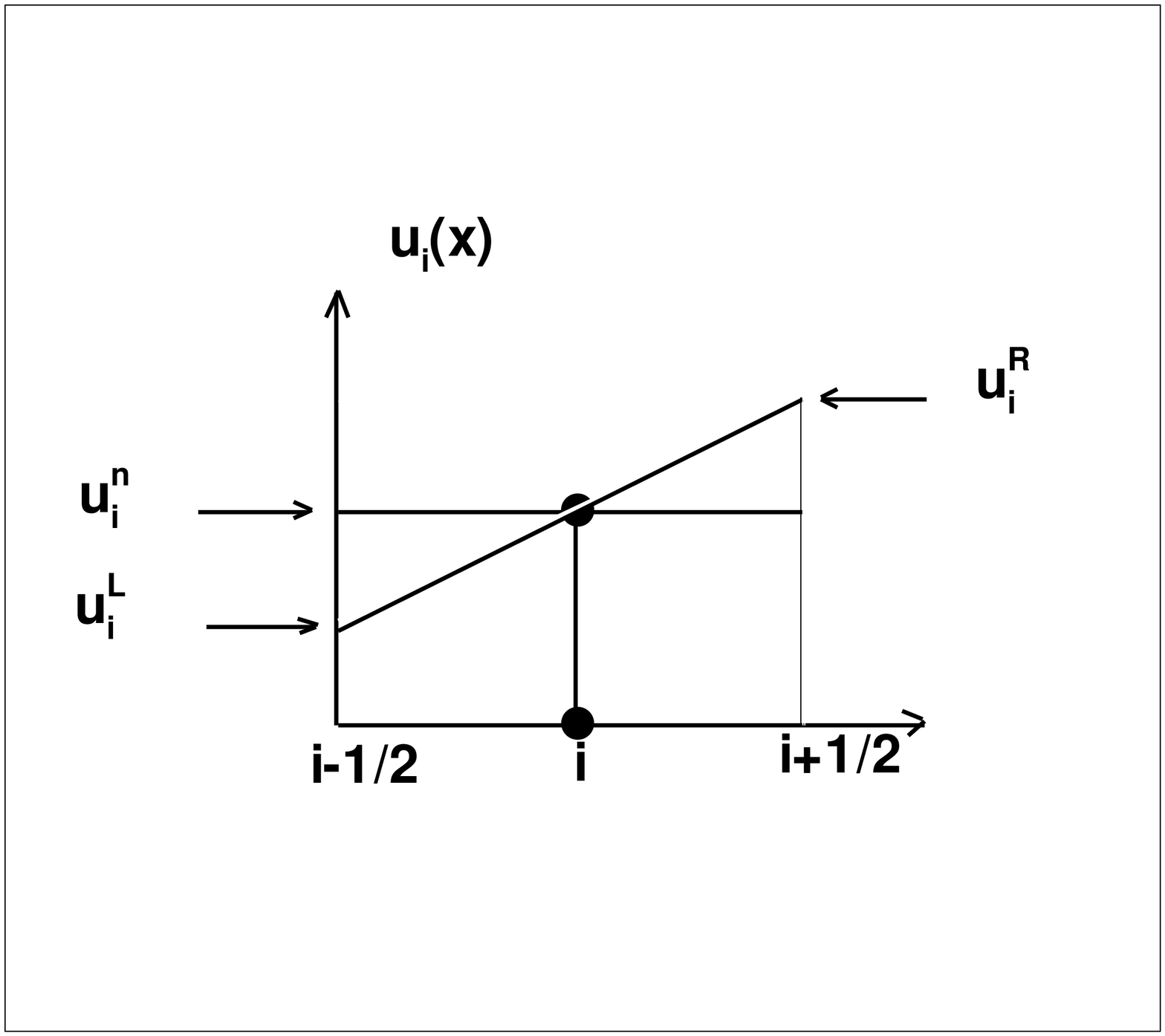}
\caption{Piecewise linear MUSCL reconstruction of a specific grid zone
$i$. The boundary extrapolated values are $u^{L}_{i}$ and  $u^{R}_{i}$}
\label{MUSCL_reconstruction}
\end{figure}
\end{center}

\noindent
Local reconstruction of $u_{i}(x)$ from
Fig.\ref{MUSCL_reconstruction} is  

\begin{eqnarray}
u_{i}(x) = u^{n}_{i} + \frac{(x - x_{i})}{\Delta x} \Delta_i,
\;\;\;\;\;\;\;\;  x \; \epsilon \; [0, \Delta x].
\label{local reconstruction}
\end{eqnarray}

\noindent
where $\frac{\Delta_i}{\Delta x}$ is called the slope  of
$u_{i}(x)$ in cell $i$.
Fig.\ref{MUSCL_reconstruction} shows the specific grid cell $i$. The
center of the cell $x_i$ in local coordinates is $x = \frac{1}{2}
\Delta x$ and $ u_{i}(x_i) = u^{n}_{i}$. From Eq.(\ref{local
reconstruction}), the values of  $u_{i}(x_i)$  at the left and right
edges of the cell play an important role in this reconstruction scheme;
they are given by 

\begin{eqnarray}
u^{L}_{i} = u_{i}(0) = u_{i} - \frac{1}{2} \Delta_i \nonumber \\
u^{R}_{i} = u_{i}(\Delta x) = u_{i} + \frac{1}{2} \Delta_i .
\label{left and right MUSCL}
\end{eqnarray}

\noindent
These left and right states are called boundary extrapolated
values. Note that the integral of $u_{i}(x)$ in cell $i$ is identical
to that of $u^{n}_{i}$ and thus the reconstruction process retains
flux conservation. This is a second-order accurate scheme,
$O(\bigtriangleup x^2)$.

If we assume the slopes are zero in Eq.(\ref{left and right MUSCL}), the
MUSCL scheme becomes the first-order accurate Godunov method.


\subsection{Slope Functions}
\label{Slope Functions}

To avoid the appearance of oscillations around discontinuities in
MUSCL-type schemes, we will use slope limiters in the reconstruction
stage \cite{HIRSCH,Toro}.

\begin{center}
\begin{figure}[ht]
\epsfxsize=15.cm \epsfysize=15.cm
\epsfbox{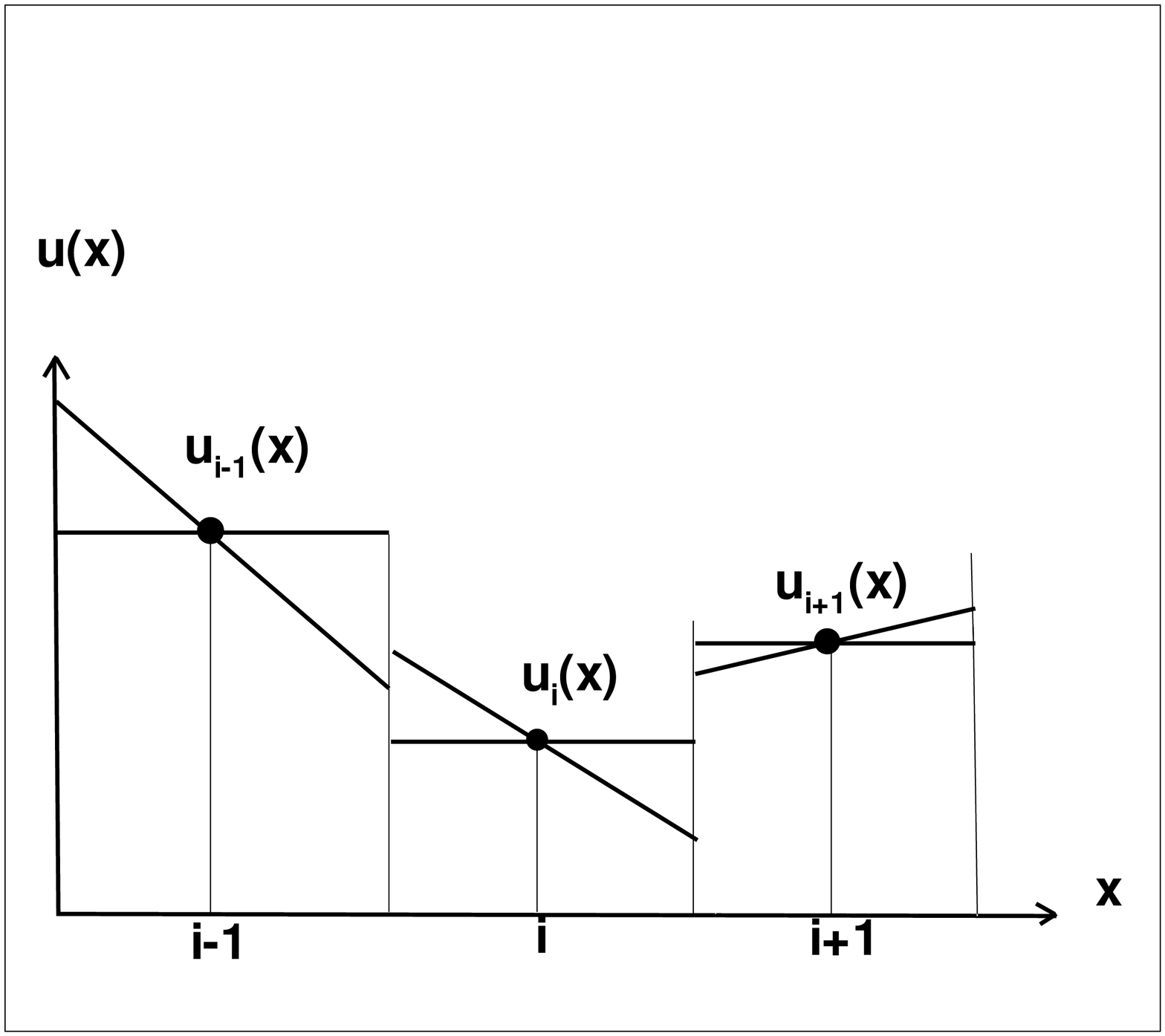}
\caption{Piecewise linear MUSCL reconstruction for three successive 
zones of $i-1$, $i$, $i+1$.}
\label{MUSCL_reconstruction2}
\end{figure}
\end{center}

Fig.\ref{MUSCL_reconstruction2} shows the piecewise linear
reconstruction process applied to three successive cells. In each
cell, we use the slope function defined in Eq.(\ref{local
reconstruction}) and (\ref{left and right MUSCL}). We will begin by writing
the slope function in the form \cite{HIRSCH}

\begin{eqnarray}
\Delta_i = \frac{1}{2}(1 + \omega)\Delta u_{i-\frac{1}{2}} +
\frac{1}{2}(1 - \omega)\Delta u_{i+\frac{1}{2}}
\label{slope function}
\end{eqnarray} 

\noindent
where 
\begin{eqnarray}
\Delta u_{i-\frac{1}{2}} \equiv u_{i}^{n} - u_{i-1}^{n}, \nonumber \\
\Delta u_{i+\frac{1}{2}} \equiv u_{i+1}^{n} - u_{i}^{n},
\label{delta operators}
\end{eqnarray}

\noindent
and $\omega$ is a free parameter in the interval $[-1, 1]$. This
produces second-order accurate schemes. For
$\omega=0$, $\Delta_i$ is a central difference approximation,
multiplied by $\Delta x$. For $\omega=-1$, the MUSCL scheme becomes
the \emph{Lax-Wendroff Method}. 

In general, schemes based on Eq.(\ref{slope function}) still have
spurious oscillations at discontinuities. To remove these, we will use
limiters that produce schemes which are total variation
diminishing, or TVD. A numerical scheme is said to be  TVD if

\begin{equation}
TV(U^{n+1}) \leq TV(U^{n}),
\label{TVD scheme}
\end{equation}

\noindent
where the total variation
\begin{equation}
TV(U^{n}) = \sum_i \mid U_{i+1} - U_i \mid.
\label{TVD scheme1}
\end{equation}

\noindent
and $i$ $\rightarrow$ $[-\infty$ , $\infty]$. To apply this rule for any
finite number of 
points on a grid, $U_i$ can be set to \emph{zero} or a \emph{constant}
value outside the grid.

A common TVD limiter is based on the minmod function
\cite{HIRSCH}. The standard minmod slope
provides the desired second-order accuracy for smooth solutions,
while still satisfying the $TVD$ property. We write this as referring
to Fig.\ref{MUSCL_reconstruction2},

\begin{eqnarray}
\Delta_i = \rm{minmod}(U_i - U_{i-1}, U_{i+1} - U_i),
\label{minmod1}
\end{eqnarray}

\noindent
where the minmod function of two arguments is defined by:

\begin{eqnarray}
 \rm{minmod}(a,b) = \left\{ \begin{array}{lll}
a & \mbox{if $|a| < |b|$ and $ab>0$} \\
b & \mbox{if $|b| < |a|$ and $ab>0$} \\
0 & \mbox{if $ab \leq 0$} .
		\end{array}
	\right. 
\label{minmod2}
\end{eqnarray}

We have also used another TVD slope limiter which may give
better solution at 
discontinuities. This limiter is given by \cite{HIRSCH}

\begin{eqnarray}
 \Delta_i = \left\{ \begin{array}{lll}
\rm{max} [0,\rm{min}(\beta
\Delta U_{i-\frac{1}{2}},\Delta U_{i+\frac{1}{2}}),\rm{min}(\Delta
U_{i-\frac{1}{2}},\beta \Delta U_{i+\frac{1}{2}})], \;\;\; \Delta
U_{i+\frac{1}{2}} > 0.0 \\   
\rm{min} [0,\rm{max}(\beta
\Delta U_{i-\frac{1}{2}},\Delta U_{i+\frac{1}{2}}),\rm{max}(\Delta
U_{i-\frac{1}{2}},\beta \Delta U_{i+\frac{1}{2}})], \;\;\; \Delta
U_{i+\frac{1}{2}} < 0.0,

		\end{array}
	\right. 
\label{slopefunction toro}
\end{eqnarray}

\noindent
where $ 1 \leq
\beta \leq 2$. The value $\beta = 1$ reproduces the 
\emph{MINMOD} or \emph{MINBEE} slope limiter as in Eq.(\ref{minmod2}). $\beta = 2$ is called
the \emph{SUPERBEE} flux limiter.


\subsection{Marquina Fluxes}
\label{Marquina Fluxes}

Approximate Riemann solver failures and their respective corrections
(usually adding a  artificial dissipation) have been studied in the
literature \cite{Quirk}. Motivated by the search for a robust and accurate
approximate Riemann solver that avoids these common failures, Shu et
al \cite{ShuOsher} have proposed a numerical flux formula for scalar
equations. Marquina flux is generalization of flux formula in
Ref. \cite{ShuOsher}. In the scalar case and for
characteristic wave speeds which do not change sign at the given
numerical interface, Marquina's flux formula is identical to Roe's
flux \cite{Toro}.  Otherwise, scheme is more viscous, entropy
satisfying local Lax-Friedrichs scheme \cite{ShuOsher}. The
combination of Roe and Lax-Friedrichs schemes is carried out in each
characteristic field after the local linearization and decoupling of
the system of equations. However, contrary to other schemes, the
Marquina's method is not based on any averaged intermediate state.

We use  Marquina fluxes with MUSCL left and right states
to solve the 
1-D relativistic hydro equation. In Marquina's scheme there are no
Riemann solutions 
involved (exact or approximate) and there are no artificial
intermediate states constructed at each cell interface. 

To compute the Marquina fluxes we first compute  the sided
local characteristic variables and fluxes. For the left and right sides,
the characteristic variables are

\begin{equation}
w_{l}^{p} = L^{p}(U_{l}) \cdot U_l, \;\;\;\; w_{r}^{p} = L^{p}(U_{r}) \cdot U_r
\label{wl} 
\end{equation}

\noindent
and the characteristic fluxes are

\begin{equation}
\Phi_{l}^{p} =  L^{p}(U_{l}) \cdot F(U_l), \;\;\;\;\; \Phi_{r}^{p} =
L^{p}(U_{r}) \cdot F(U_r). 
\end{equation}

\noindent
where the number of conservative variables $p = 1..5$. $U_l$ and $U_r$
are conservative variables at the left and 
right sides, respectively. $L^{p}(U_{l})$ and $L^{p}(U_{r})$ are the left 
eigenvectors of the Jacobian matrices, $\partial F^i/\partial U$.

We define left and right fluxes depending on the velocities of the fluid for
each specific grid zone. The prescription given in Ref.\cite{DFIM} is
as follows.

For all conserved variables $p = 1, ..m$

if \emph{$\lambda_{p}(U)$} does not change sign in (if
( \emph{$\lambda_{p}(U_L)$ $\times$ $\lambda_{p}(U_R)$ $\geq$ 0))}, then 

$\;\;\;\;$ if \emph{$\lambda_{p}(U_l) > 0$} then

$\;\;\;\;\;\;\;\;\;\;\;$	$\Phi_{+}^{p} = \Phi_{l}^{p}$
	
$\;\;\;\;\;\;\;\;\;\;\;$	$\Phi_{-}^{p} = 0$

$\;\;\;\;$   else
	
$\;\;\;\;\;\;\;\;\;\;\;$	$\Phi_{+}^{p} = 0$
	
$\;\;\;\;\;\;\;\;\;\;\;$	$\Phi_{-}^{p} = \Phi_{r}^{p}$	

$\;\;\;\;$ end if

else

$\;\;\;\;\;\;\;\;\;\;\;$   $\alpha_{p} =  \max_{U \epsilon \Gamma(U_l,
U_r)} | \lambda_p(U) | $

$\;\;\;\;\;\;\;\;\;\;\;$ $\Phi_{+}^{p} = 0.5 ( \Phi_{l}^{p} +
\alpha_{k} w_{l}^{p})$

$\;\;\;\;\;\;\;\;\;\;\;$ $\Phi_{-}^{p} = 0.5 ( \Phi_{r}^{p} +
\alpha_{p} w_{r}^{p})$ 

end if

\noindent
where $\lambda_p$ is an eigenvalue of the Jacobian matrix and,
\begin{eqnarray}
\alpha_k = \max \{ |\lambda_{p} (U_l)|, |\lambda_{p} (U_r)| \}.
\label{alpha k}
\end{eqnarray}

\noindent
The numerical flux that corresponds to the cell interface
separating the states $U_l$ and $U_r$ is then given by Ref.\cite{DFIM}:

\begin{eqnarray}
F^{M} (U_l, U_r) = \sum_{p=1}^{m} ( \Phi_{+}^p r^p (U_l) +
\Phi_{-}^p r^p (U_r)  ).
\label{marquina flux}
\end{eqnarray}

Marquina's scheme can be  interpreted as a characteristic-based
scheme that avoids the use of an averaged intermediate state to
perform the transformation to the local characteristic fields. 

In carrying out Marquina's scheme, we have to compute intermediate
states and the Jacobian matrix of the states at each cell interface. So we
need to know the left and right states, $U_L$ and  $U_R$, at each
interface. To construct the 
second-order scheme, we use the MUSCL left and right states  given
in Eq.(\ref{left and right MUSCL}).


\section{NUMERICAL RESULTS}
\label{Numerical Results}

Results of numerical solution of SRH equation are given. Before doing any further 
explanation, we need to define boundary conditions. Boundary conditions are set 
by filling the data in guard cells  with
appropriate values. In the numerical calculation boundary filling
plays an important role in the simulations. The computational grid is
extended at both sides of the physical domain to compute the fluxes at
interfaces. These extra cells are also called guard cells or ghost
zones. There are different 
types of boundary conditions used in the literature to solve physical
problems in an appropriate way. In this paper we have used several
types of boundary conditions including periodic, inflow, outflow and
analytically prescribed boundary conditions. These boundary conditions
have to be provided on each time step for all primitive and
conservative variables in the special relativistic hydro code.

Here, we solve three different test problems to compare the results from different 
numerical schemes.


\subsection{Smooth Test Problems}
\label{Smooth Test Problem}

First, we start testing the code with smooth 
hydrodynamical solutions using
different numerical schemes which are explained in \ref{Numerical Schems}. Since
we are concerned with special relativistic flows, we choose cases with
$P \sim \rho$ and $v \sim 1$ in our units ($c =1$). We focus on the
case of a varying density profile $\rho = \rho (x,t)$ with constant,
uniform pressure $P = P_0$ and velocity $v = v_0$. When these
functions are substituted into Eq. (\ref{desired equation}), we
see that they form a consistent solution for the advection of a
density profile at constant velocity $v_0$. These tests are performed
on the computational domain $0 \leq x \leq 1$ with the ideal gas law 
Eq.(\ref{flux split2}), with $\Gamma = 5/3$.

\begin{center}
\begin{figure}
\centerline{\epsfxsize=15cm \epsfysize=15cm
\epsffile{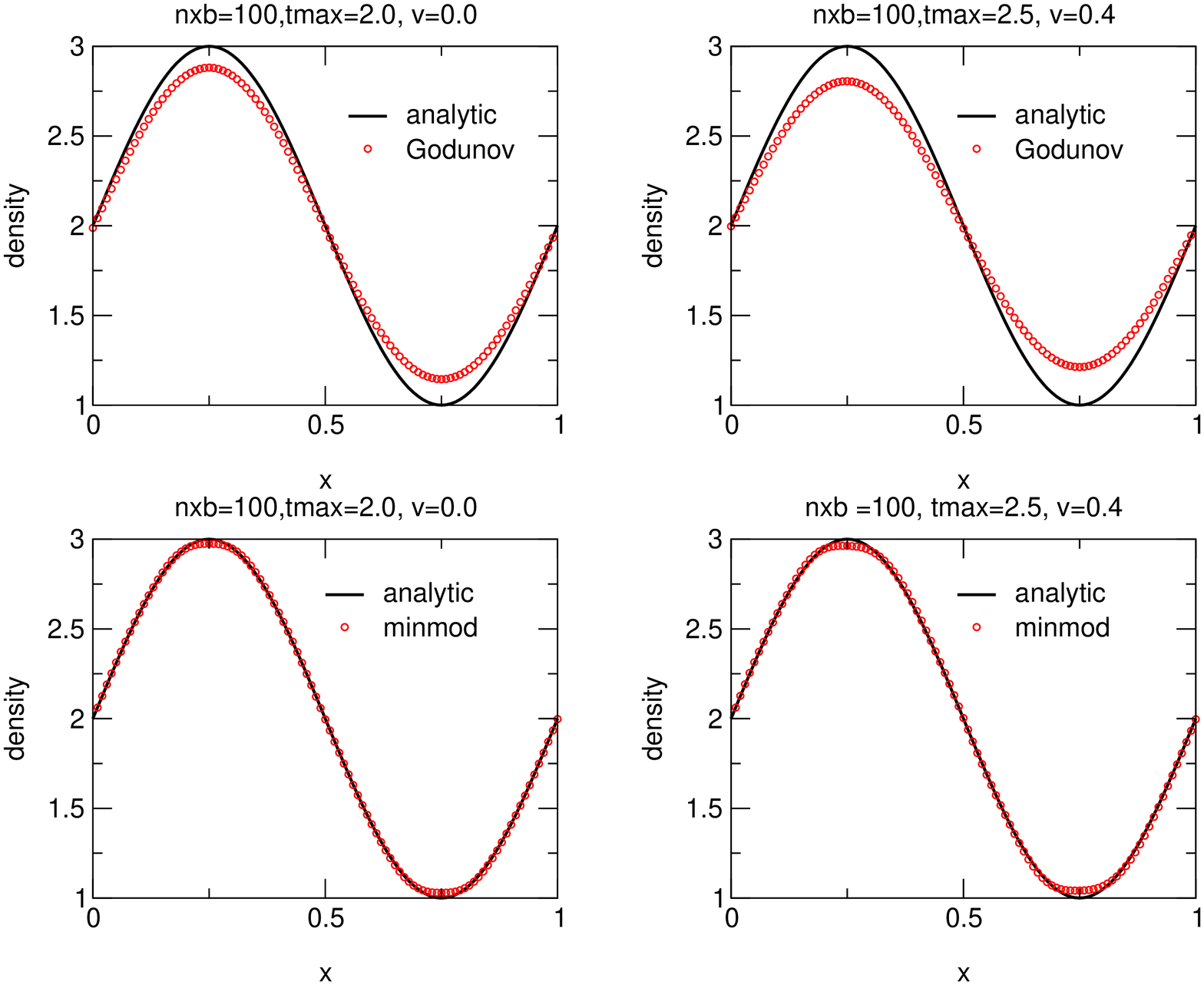}}
\caption{Plot for standing and moving waves using the Godunov method
and the MUSCL scheme with the minmod limiter Eq.(\ref{minmod2}). $npts = 100$.}
\label{Mov-stand_differnt schemes 1}
\end{figure}
\end{center}

\begin{center}
\begin{figure}
\centerline{\epsfxsize=15cm \epsfysize=15cm
\epsffile{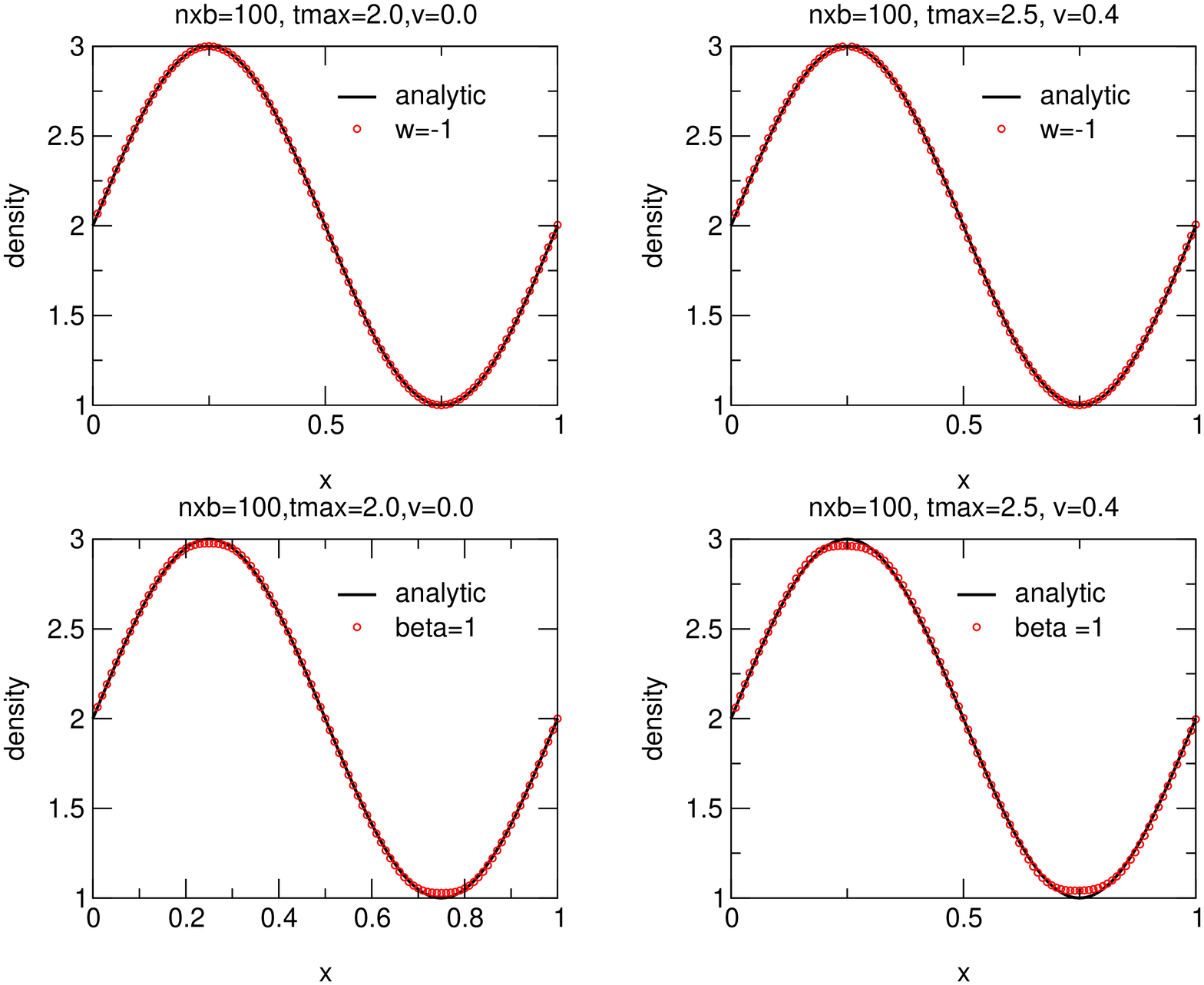}}
\caption{Plot for standing and moving waves using the  slopes functions
$w=-1$ in Eq.(\ref{slope function})(Lax-Wendroff scheme) and  $\beta
= 1$ in Eq.(\ref{slopefunction toro}). $npts = 100$.}
\label{Mov-stand_differnt schemes 2}
\end{figure}
\end{center}

\begin{center}
\begin{figure}
\centerline{\epsfxsize=15cm \epsfysize=15cm
\epsffile{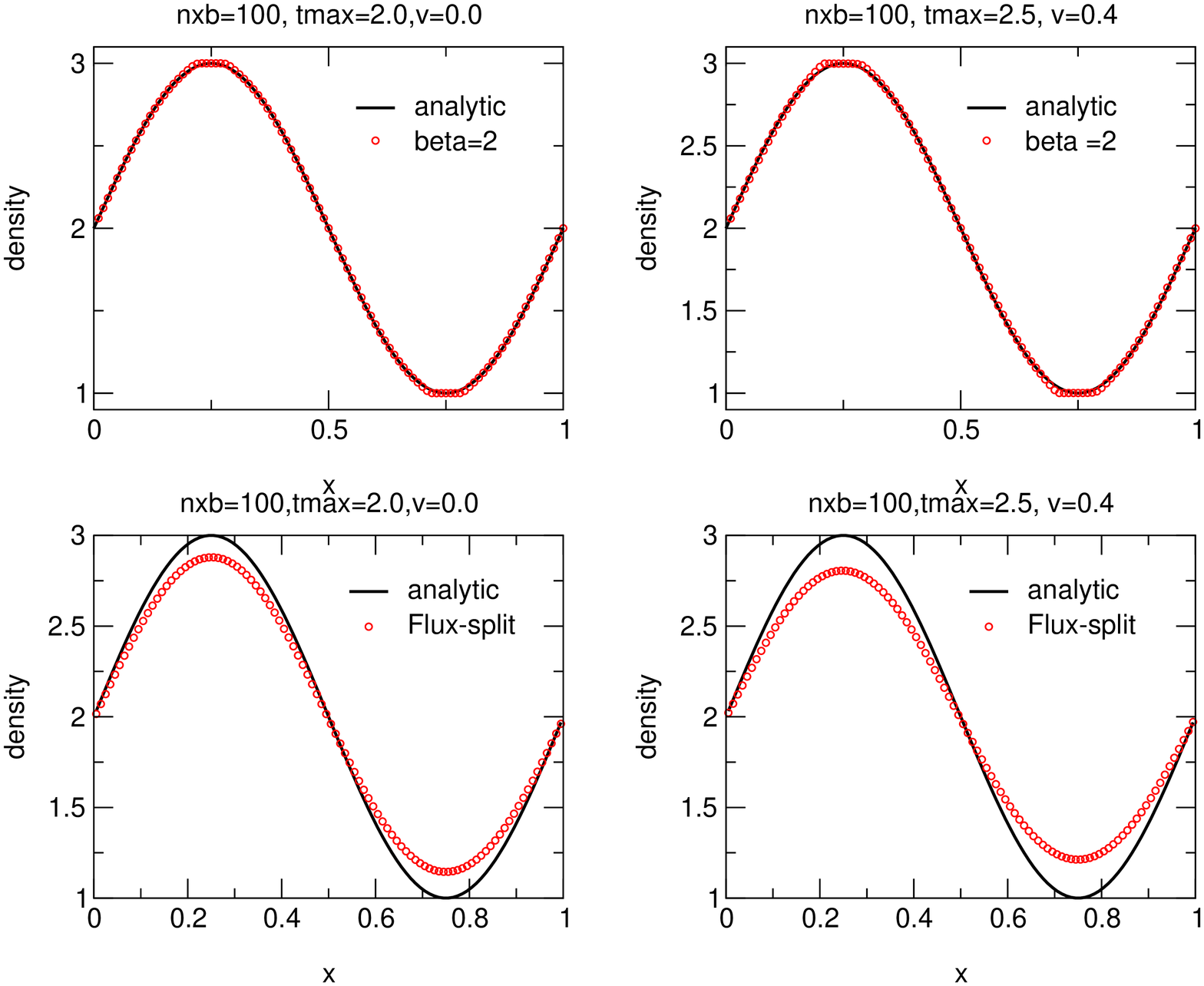}}
\caption{Plot for standing and moving waves using  the slope function
for$\beta = 2$ in Eq.(\ref{slopefunction toro}) and the flux splitting
method. $npts = 100$.} 
\label{Mov-stand_differnt schemes 3}
\end{figure}
\end{center}

The first test in Table \ref{Special Relativistic smooth test problem}
consists of a stationary density pulse.In Table \ref{Comparing L1 errores 
for standing wave}  we compute the $L_1$ norm errors and convergence rates, 
$c$, for the different numerical scheme for the standing wave test problem in 
Table.\ref{Special Relativistic smooth test problem}.  All numerical 
schemes give a good convergence
rate for the standing wave problem, except the minmod schemes. However
 while $TVD$ schemes completely remove  the 
oscillations, they reduce locally the accuracy of the solution around
the extrema. We also compare the numerical solutions of the standing
wave, shown in the left-hand panels and  labeled with $v=0$, with the 
analytic solutions using these
schemes in Figs. \ref{Mov-stand_differnt schemes 1}, \ref{Mov-stand_differnt
schemes 2} and \ref{Mov-stand_differnt schemes 3}. It is easy to see
from these figures that the $TVD$ schemes($minmod$, $\beta = 1$ and $\beta =
2$) reduce the accuracy of the solution around the extrema. From
Fig. \ref{Mov-stand_differnt schemes 2} with $w=-1$, the
Lax-Wendroff scheme gives better solution for the smooth wave.

In Table \ref{Comparing L1 errores for moving wave} we compute the
$L_1$ norm errors and convergence rates,$c$, using the different numerical
schemes for the moving wave in Table \ref{Special Relativistic
smooth test problem}. We got good first-order convergence rates for
the flux splitting and Godunov methods. The Lax-Wendroff method gives
good convergence rates for second-order method. The convergence rates
with $TVD$ schemes are not as good as for Lax-Wendroff, and they are not
consistent because of the problems around the extrema.
In Figs. \ref{Mov-stand_differnt schemes 1},
\ref{Mov-stand_differnt schemes 2} and \ref{Mov-stand_differnt schemes
3}, we plot the numerical solutions of the moving wave, shown in the 
right-hand panels and  label with $v=0.4$, with the
analytic solutions using different schemes. Again, the $TVD$ schemes
reduce the accuracy of the solution around the extrema. 


\subsection{Shock Tube Test Problem}
\label{Shock Test Problem}

Our next code test is the Riemann shock tube \cite{HIRSCH,Toro}.  
In this problem, the fluid is initially in two
different thermodynamical states on either side of a membrane. The
membrane is then removed. Let us assume that the fluid initially has
$\rho_L > \rho_R$, where the subscripts $L$ and $R$ refer to the left
and right sides of the membrane. Then, a rarefaction wave travels to
the left, and a shock wave and contact discontinuity travel to the
right.The Riemann shock tube
is a useful test problem because it has an exact time-dependent
solution and tests the ability of the code to evolve both smooth and
discontinuous flows. In the case considered here, the velocities are
special relativistic and the method of finding the exact solution
differs somewhat from the standard non-relativistic shock tube. 

\begin{center}
\begin{figure}
\centerline{\epsfxsize=15cm \epsfysize=15cm
\epsffile{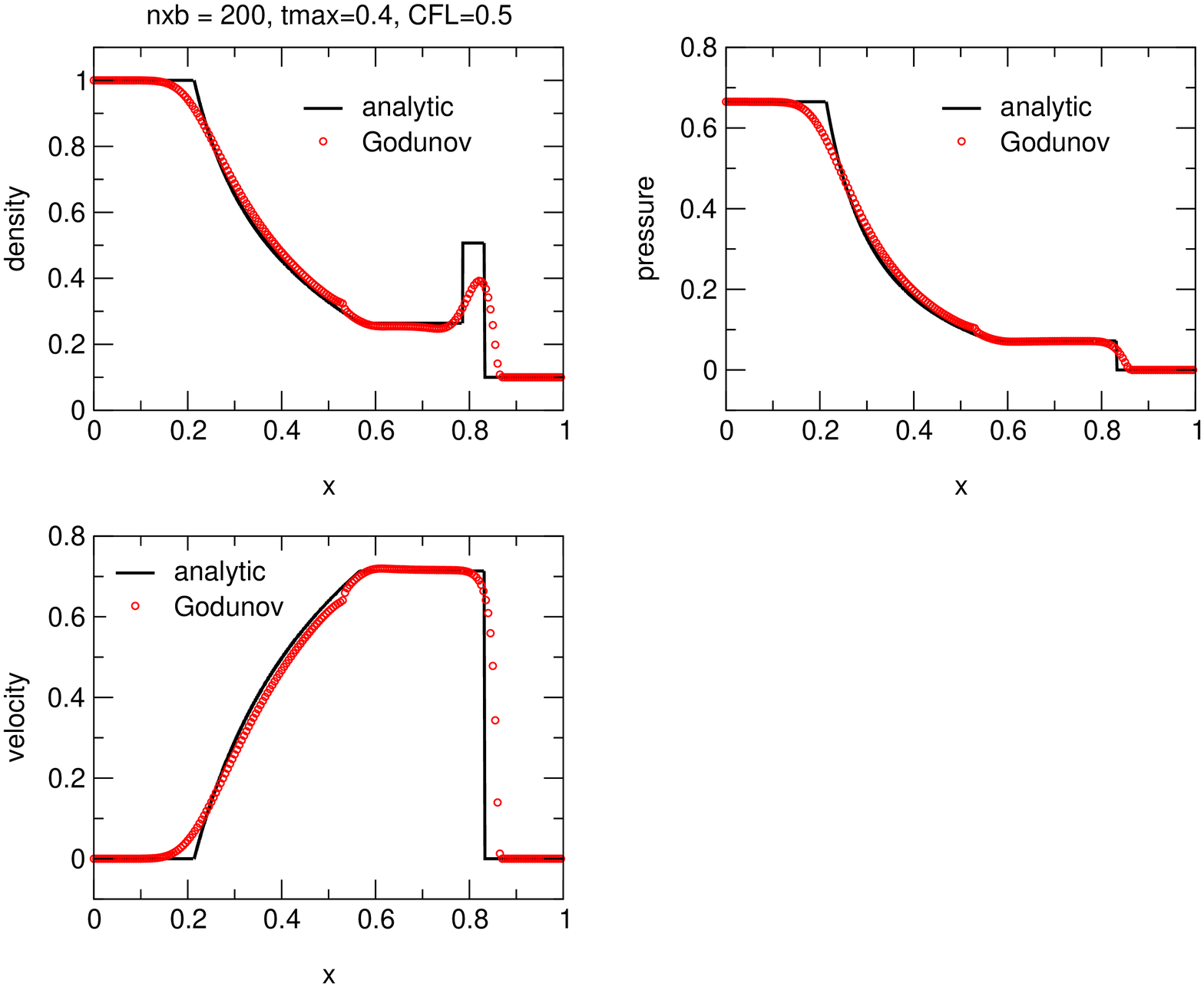}}
\caption{The analytic and numerical solutions of the relativistic
shock tube problem are  plotted. The first-order 
Godunov scheme is used with resolution \; $npts = 100$.}
\label{compare shock tube 1}
\end{figure}
\end{center}

\begin{center}
\begin{figure}
\centerline{\epsfxsize=15cm \epsfysize=15cm
\epsffile{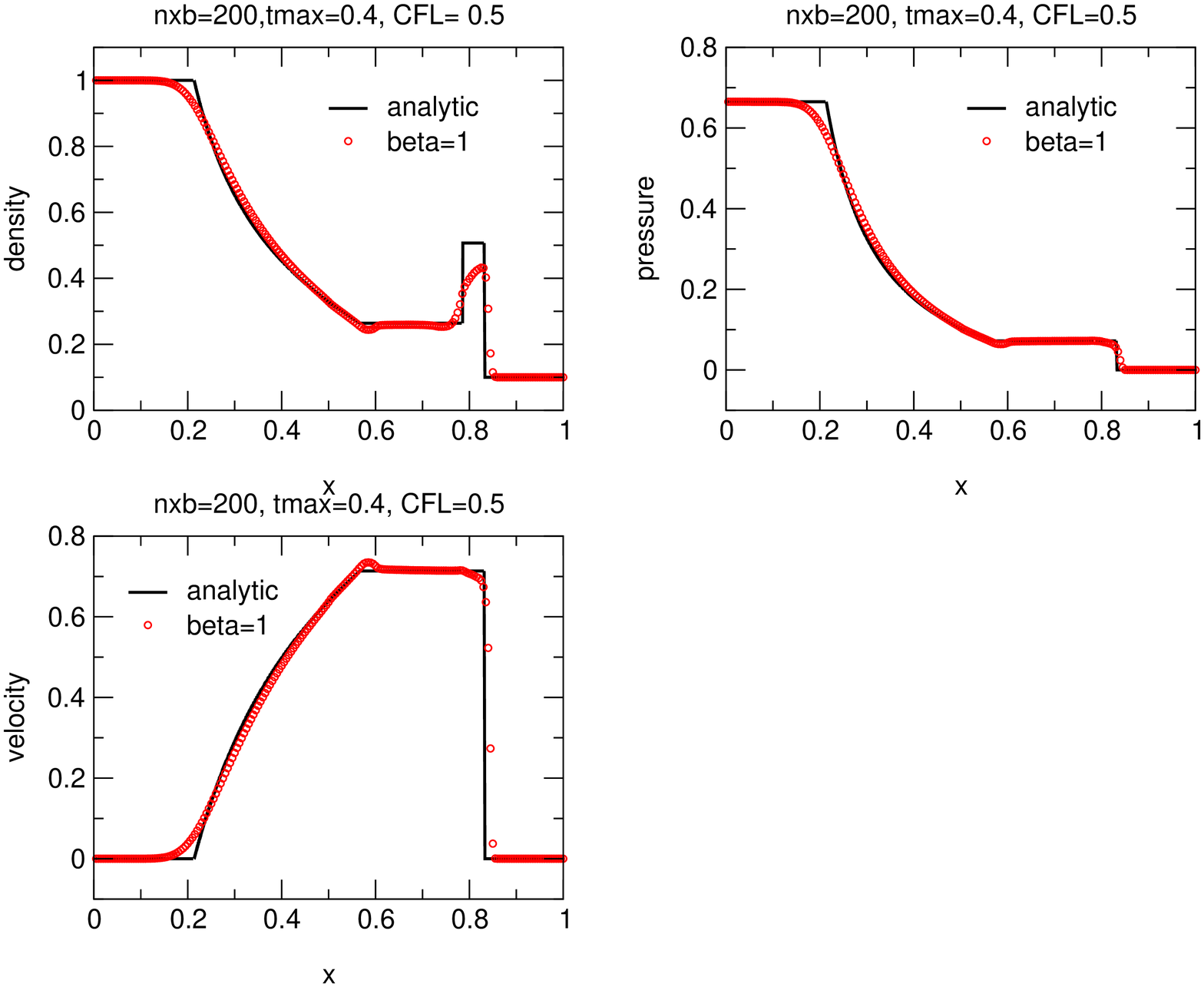}}
\caption{The analytic and numerical solutions of the relativistic
shock tube problem are plotted. The slope function for $\beta = 1$ in
Eq.(\ref{slopefunction toro}) is used with  resolution $npts = 100$.}
\label{compare shock tube 2}
\end{figure}
\end{center}

\begin{center}
\begin{figure}
\centerline{\epsfxsize=15cm \epsfysize=15cm
\epsffile{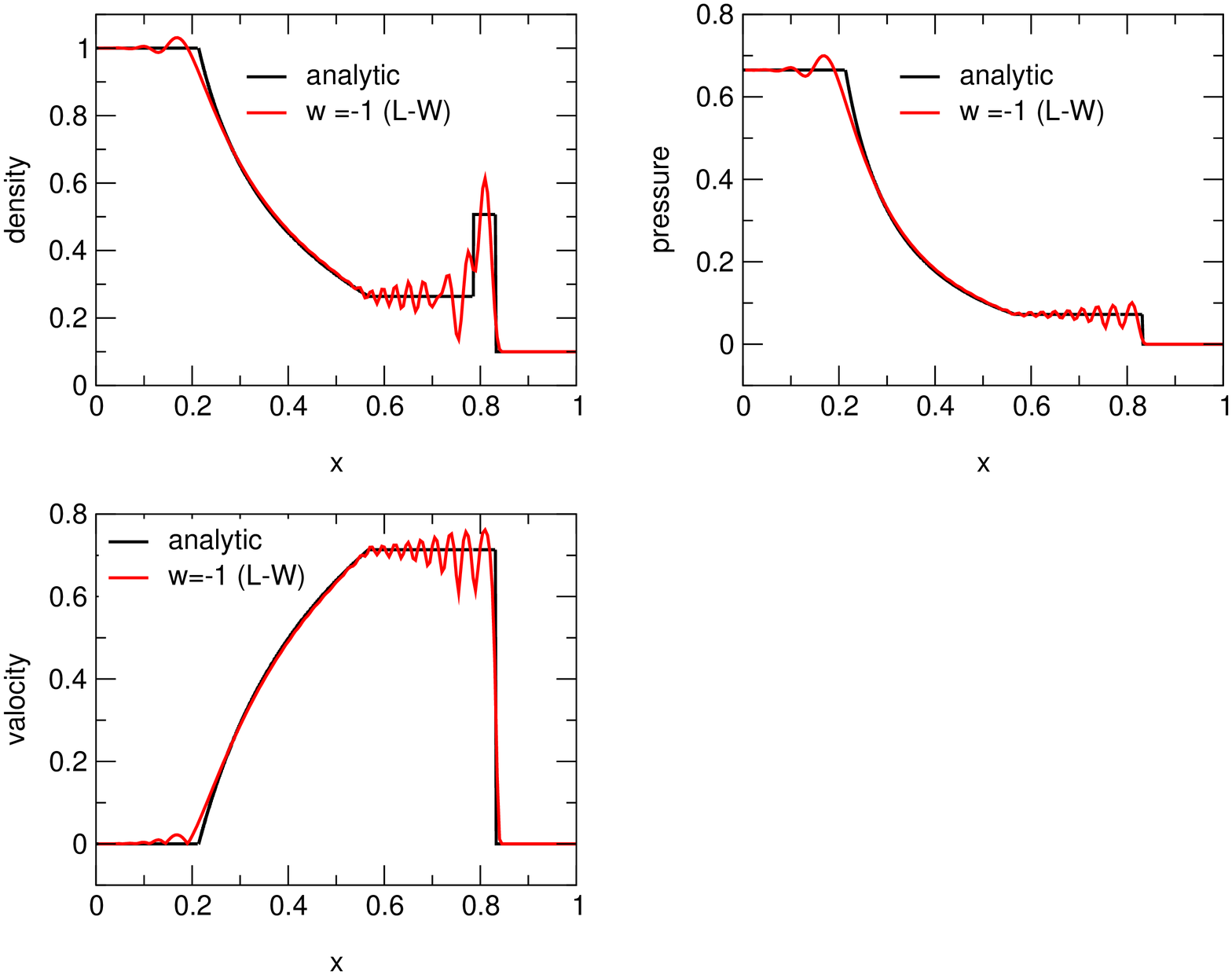}}
\caption{The analytic and numerical solutions of the relativistic
shock tube problem are plotted. The slope function for  $w=-1$ in
Eq.(\ref{slope function})(Lax-Wendroff scheme) is used with 
resolution $npts = 100$.} 
\label{compare shock tube 3}
\end{figure}
\end{center}

In Table \ref{Comparing L1 errores for shock wave} we compute the 
$L_1$ norm errors and convergence rates using the different numerical
schemes for the special relativistic shock problem in Table 
\ref{special relativistic shock tube test problems}. The  convergence rates 
should approach $1$ when we use higher order
methods. From the last three columns of Table \ref{Comparing L1 
errores for shock wave},  the first-order flux splitting  and Godunov
methods give good convergence rates, but not the
Lax-Wendroff scheme, which scheme produces spurious
oscillation behind the shock. This is  seen clearly in
Fig.\ref{compare shock tube 3}. The $TVD$ schemes give good
convergence rates for the shock tube problem. From Table
\ref{Comparing L1 errores for shock wave} the $TVD$ schemes give better
convergence rates than the flux-splitting and Godunov schemes, because $TVD$
schemes are second-order accurate. Additionally, we plot the
analytic and numerical solutions of the shock tube problem for Godunov and
$TVD$ with $\beta=1$ in Figs. \ref{compare shock tube 1} and
\ref{compare shock tube 2}. We did not compute the convergence rates for  $\beta=2$. 
Because it produce some oscillation and it does not allow to 
us run the code enough time to compute convergence rates.

\section{CONCLUSION}

Numerical solution of special relativistic hydrodynamical equation in $1D$ using 
first and second order different numerical methods is explained in this paper.
The numerical methods are applied on cases which are stationary and unsteady flow
situations. Results from different method are compared to define better method 
for problems. It is seen from figures and tables that while $TVD$ type schemes gives 
good approximation for discontinuity solution, the $Non-TVD$ type schemes give 
better solution for smooth test problems. Because $TVD$ schemes completely remove  
the oscillations, they reduce locally the accuracy of the solution around
the extrema. As a conclusion, $TVD$ type schemes can use  to solve astrophysical 
problems which have strong shock region, especially around the compact objects.


\begin{acknowledgments}
This project has been performed using NASA and Pittsburgh super 
computers/T3E clusters.
\end{acknowledgments}

\newpage

\begin{table}
\caption{Initial data for smooth waves test problems.} 
$$  \vbox{ \offinterlineskip \vskip7pt
  \def\qq{\hskip1.0em}
  \def\laststrut{\vrule depth6pt width0pt}
  \def\titlestrut{\vrule height12pt depth6pt width0pt}
\halign {#\vrule\strut &\quad#\hfil
       &&\qq#\vrule &\qq\hfil#\hfil \cr
\noalign{\hrule}
 & \multispan{7}\hfil Special Relativistic Smooth Wave Test Problems
  \hfil\titlestrut  
  &\cr \noalign{\hrule}  
& {\em Test} && {\em $\rho$} && {\em $P$} &&  {\em $v$}  &\cr
 \noalign{\hrule}
& 1 && $\sin(2 \pi x) + 2.0$ && 1.0 && 0.0 &\cr
 \hline
& 2 && $\sin(2 \pi x) + 2.0$ && 1.0 && 0.4  \laststrut &\cr
\noalign{\hrule\vskip4pt}
  \noalign{\vskip5pt} }}$$
\label{Special Relativistic smooth test problem}
\end{table}

\newpage

\begin{table}
\caption{$L_1$ norm errors and convergence rates for the standing wave
test problem in Table 
\ref{Special Relativistic smooth test problem}. The different first
and second-order schemes are used.}
  \vbox{ \offinterlineskip \vskip5pt
  \def\qq{\hskip0.6em}
  \def\laststrut{\vrule depth4pt width0pt}
  \def\titlestrut{\vrule height11pt depth4pt width0pt}
\halign {#\vrule\strut &\quad#\hfil
       &&\qq#\vrule &\qq\hfil#\hfil \cr
\noalign{\hrule}
 & \multispan{15}\hfil L1 norm errors and convergence rates for
  the standing wave \hfil\titlestrut  
  &\cr \noalign{\hrule}  
& Type && $npts$&& $L_1(\rho)$&& $L_1(p)$&& $L_1(v)$ && $c(\rho)$ &&
  $c(p)$ && $c(v)$ &\cr 
  \noalign{\hrule}
&          && 100 && 8.42E-2 &&  1.56E-3 &&  7.84E-4 && 1.92 && 1.88  &&  1.84 &\cr 
& Flux-splitting && 200 && 4.36E-2 && 8.29E-4 && 4.25E-4  && 1.96 && 1.94 && 1.92  &\cr
& $O(\Delta x,\Delta t)$ && 400 && 2.22E-2 && 4.28E-4 && 2.21E-4 &&
1.98 && 1.97 && 1.95 \laststrut &\cr  
& (non-TVD)&& 800 && 1.12 E-2 && 2.16E-4 && 1.13E-4 && 1.98 && 1.99 && 1.96   &\cr 
&          && 1600 && 5.64 E-3 && 1.09E-4 && 5.77E-5 && && &&  \laststrut &\cr
  \noalign{\hrule}
&          && 100 && 8.38E-2 && 1.87E-3 && 5.67E-4 && 1.93 &&  1.84&&
1.82 &\cr 
& Godunov  && 200 && 4.33E-2 && 1.01E-3 && 3.10E-4 && 1.96 && 1.92 &&
1.90 &\cr 
& $O(\Delta x,\Delta t)$ && 400 && 2.20E-2 && 5.27E-4 && 1.62E-4 &&
1.98 && 1.95 && 1.95 \laststrut &\cr  
& (non-TVD)&& 800 && 1.11E-2 && 2.69E-4 && 8.33E-5 && 1.99  && 1.97 && 1.97 &\cr
&          && 1600 && 5.59E-3  && 1.36E-4 && 4.22E-5 && && && \laststrut &\cr 
  \noalign{\hrule}
& w=-1      && 100 && 2.87E-3 && 7.06E-5 && 2.56E-5 && 3.99 && 3.99 &&
4.00 &\cr 
& (Lax-Wend.)&& 200 && 7.19E-4 && 1.76E-5 && 6.39E-6 && 3.99 &&
3.999  && 4.00 &\cr 
& $O(\Delta x^2,\Delta t^2)$ && 400 && 1.79E-4 && 4.41E-6 &&  1.59E-6
&& 3.999 && 3.999 && 3.91 \laststrut &\cr  
& (non-TVD)&& 800 && 4.49E-5 && 1.10E-6 && 4.07E-7 && && && \laststrut &\cr
  \noalign{\hrule}
&          && 100 && 6.55E-3 && 4.37E-5 && 2.49E-5 && 3.67 &&3.58  &&
3.67 &\cr
& $\beta =1$ && 200 && 1.78E-3 && 1.22E-5 && 6.79E-6 && 3.65  && 3.68
&& 3.68 &\cr 
& $O(\Delta x^2,\Delta t^2)$ && 400 && 4.87E-4 && 3.32E-6 && 1.84E-6 &&
3.76 &&  3.79&& 3.77 \laststrut &\cr  
& (TVD)    && 800 && 1.29E-4 && 8.75E-7 && 4.89E-7 && 3.81 && 3.87 && 3.84 &\cr
&          && 1600 && 3.39E-5 && 2.25E-7 &&  1.27E-7 && && &&  \laststrut &\cr
  \noalign{\hrule}
&          && 100 && 4.69 E-3&& 3.75E-5 && 1.59E-5 && 3.53 && 3.32 &&
2.96 &\cr 
& $\beta =2$ && 200 &&1.32E-3  && 1.13E-5 && 5.37 E-6 && 3.77  && 3.53
&& 3.36 &\cr
& $O(\Delta x^2,\Delta t^2)$ && 400 && 3.51E-4 && 3.19E-6 && 1.59E-6 &&
3.88 && 3.72  && 3.56 \laststrut &\cr  
& (TVD)   && 800 && 9.04 E-5 && 8.58E-7 && 4.48E-7 && 3.94 && 3.84  &&
3.72  &\cr 
&          && 1600 && 2.29 E-5 && 2.23E-7 && 1.20E-7 && && &&   \laststrut &\cr
  \noalign{\hrule}
&          && 100 && 6.38E-3 &&  5.47E-4 &&  1.68E-4 && 3.56 && 1.80
&& 1.64 &\cr 
& minmod   && 200 && 1.79E-3 && 3.03E-4 && 1.02E-4 && 3.14 && 2.95 &&
2.32 &\cr 
& $O(\Delta x^2,\Delta t^2)$ && 400 &&5.69E-4  &&1.02E-4  && 4.42E-5 &&
3.22  && 2.89 && 3.26 \laststrut &\cr  
& (TVD)    && 800 &&1.76E-4  &&3.55E-5  && 1.35E-5 &&  2.36  && 1.32  && 1.33 &\cr
&          && 1600 &&7.45 E-5 && 2.68E-5  && 1.01E-5 && && &&  \laststrut &\cr
\noalign{\hrule\vskip4pt}
  \noalign{\vskip5pt} }} 
\label{Comparing L1 errores for standing wave}
\end{table} 

\newpage
\begin{table}
\caption{$L_1$ norm errors and convergence rates for the moving wave
test problem from Table 
\ref{Special Relativistic smooth test problem}. The different 
first and second order schemes are used.}
 \vbox{ \offinterlineskip \vskip5pt
  \def\qq{\hskip0.6em}
  \def\laststrut{\vrule depth6pt width0pt}
  \def\titlestrut{\vrule height12pt depth6pt width0pt}
\halign {#\vrule\strut &\quad#\hfil
       &&\qq#\vrule &\qq\hfil#\hfil \cr
\noalign{\hrule}
 & \multispan{15}\hfil L1 norm errors and convergence rates for the
  moving wave \hfil\titlestrut 
  &\cr \noalign{\hrule}  
& Type && $npts$&& $L_1(\rho)$&& $L_1(p)$&& $L_1(v)$ && $c(\rho)$ &&
  $c(p)$ && $c(v)$ &\cr
  \noalign{\hrule}

&          && 100 && 0.129 && 2.76E-3 && 8.79E-4 && 1.89 && 1.76  &&
1.76 &\cr 
& Flux-splitting && 200 &&  6.83E-2 && 1.56E-3 && 4.97E-4 && 1.94 &&
1.87 && 1.87 &\cr 
& $O(\Delta x,\Delta t)$ && 400 && 3.51E-2 && 8.33E-4 && 2.65E-4 &&
1.97 && 1.93 && 1.93 \laststrut &\cr  
& (non-TVD)&& 800 && 1.78E-2 && 4.30E-4 && 1.36E-4 && 1.98 && 1.96 &&
1.96  &\cr 
&          && 1600 && 8.97E-3  && 2.18E-4 &&  6.95E-5 && && && \laststrut &\cr
  \noalign{\hrule}
&          && 100 && 0.13 && 2.8E-3 &&  8.8E-4 && 1.9 && 1.78 &&
1.78 &\cr
& Godunov  && 200 && 6.84E-2&& 1.58E-3 && 4.99E-4 && 1.95 && 1.89 &&
1.89 &\cr 
& $O(\Delta x,\Delta t)$ && 400 && 3.54E-2 && 8.35E-4 && 2.67E-4 &&
1.98 && 1.93 && 1.93  \laststrut &\cr 
& (non-TVD)&& 800 && 1.8E-2 && 4.32E-4 && 1.38E-4 && 1.99 && 1.97 && 1.97 &\cr
&          && 1600 && 8.9E-3 && 2.2E-4 && 6.9E-5 && && && \laststrut &\cr
  \noalign{\hrule}
&  w=-1    && 100 && 3.94E-3 &&  1.13E-4 &&  3.67 E-5 && 3.99 && 3.99
&&  3.99  &\cr
& (Lax-Wend.)&& 200  && 9.86E-4 && 2.83E-5 && 9.19E-6  && 4.00 && 3.99
&& 3.99 &\cr
& $O(\Delta x^2,\Delta t^2)$ && 400 && 2.46E-4 && 7.08E-6 && 2.30E-6
&& 4.00 && 4.00 && 3.99  \laststrut &\cr  
& (non-TVD)&& 800 && 6.16E-5 && 1.77E-6 && 5.75E-7 && && && \laststrut &\cr
  \noalign{\hrule}
&          && 100 &&1.15 E-2 && 8.65E-5 && 3.01E-5 && 3.56 &&  3.95 &&
4.07 &\cr 
& $\beta =1$ && 200 && 3.22E-3 && 2.19E-5 &&  7.37E-6 && 3.67 && 3.96
&& 4.05  &\cr
& $O(\Delta x^2,\Delta t^2)$ && 400 && 8.77E-4 && 5.52E-6 && 1.81E-6
&& 3.73  && 3.97 && 4.01 
\laststrut &\cr  
& (TVD)    && 800 && 2.34E-4 && 1.38E-6 && 4.53E-7 && 3.81  && 3.98 && 4.00 &\cr  
&          && 1600 && 6.15E-5 && 3.48E-7 && 1.13E-7 && && &&  \laststrut &\cr
  \noalign{\hrule}
&          && 100 && 7.81E-3 && 8.06E-5 && 2.94E-5 && 3.45 && 3.81 &&
  4.43 &\cr 
& $\beta =2$ && 200 && 2.26E-3 && 2.11E-5 && 6.65E-6 && 3.71  && 3.91
  && 3.98  &\cr
& $O(\Delta x^2,\Delta t^2)$ && 400 &&6.09E-4 && 5.39E-6 && 1.67E-6 &&
  3.85 && 3.90 && 3.81 \laststrut &\cr  
& (TVD)   && 800 &&  1.57E-4 && 1.38 E-6 && 4.38E-7 &&3.92 && 4.01  &&
3.94 &\cr 
&          && 1600 && 4.01E-5 && 3.43E-7 && 1.11 E-7 && && && \laststrut &\cr
  \noalign{\hrule}
&          && 100 &&  1.15E-2 && 8.67E-5  && 3.01E-5 && 3.56 &&  3.86
  &&  4.14  &\cr
& minmod   && 200 && 3.22E-3 && 2.24E-5 && 7.25E-6 && 3.67 &&  4.05&&
  3.98  &\cr
& $O(\Delta x^2,\Delta t^2)$ && 400 && 8.77E-4  && 5.52E-6  && 1.82E-6
  && 3.73  && 3.91  && 4.03 
\laststrut &\cr  
& (TVD)    && 800 && 2.34E-4  && 1.41E-6 &&  4.50E-7 &&  3.81 && 4.04  && 3.97   &\cr
&          && 1600 && 6.15E-5  &&  3.48E-7 && 1.13E-7 && && && \laststrut &\cr
\noalign{\hrule\vskip4pt}
  \noalign{\vskip5pt} }} 
\label{Comparing L1 errores for moving wave}
\end{table} 

\newpage
 \begin{table} 
\caption{Initial data for the special relativistic shock tube test problems}
$$  \vbox{ \offinterlineskip \vskip7pt
  \def\qq{\hskip1.0em}
  \def\laststrut{\vrule depth6pt width0pt}
  \def\titlestrut{\vrule height12pt depth6pt width0pt}
\halign {#\vrule\strut &\quad#\hfil
       &&\qq#\vrule &\qq\hfil#\hfil \cr
\noalign{\hrule}
 & \multispan{13}\hfil Special Relativistic Test Problem \hfil\titlestrut 
  &\cr \noalign{\hrule}  
& {\em Test} && {\em $\rho_L$} && {\em $u_L$} && {\em $p_L$} && {\em
  $\rho_R$} && {\em $u_R$} && {\em $p_R$} &\cr
 \noalign{\hrule}
& 1 && 10.0 && 0.0 && 13.3 && 1.0 && 0.0 && $0.66 . 10^{-6}$  \laststrut &\cr
\noalign{\hrule\vskip4pt}
  \noalign{\vskip5pt} }} $$ 
\label{special relativistic shock tube test problems}
\end{table}

\newpage
\begin{table}
\caption{$L_1$ norm errors and convergence rates for the shock wave
test problem from Table 
\ref{special relativistic shock tube test problems}. The different 
first and second order schemes are used.}
 \vbox{ \offinterlineskip \vskip5pt
  \def\qq{\hskip0.6em}
  \def\laststrut{\vrule depth6pt width0pt}
  \def\titlestrut{\vrule height12pt depth6pt width0pt}
\halign {#\vrule\strut &\quad#\hfil
       &&\qq#\vrule &\qq\hfil#\hfil \cr
\noalign{\hrule}
 & \multispan{15}\hfil L1 norm errors and convergence rates for the
  shock wave \hfil\titlestrut 
  &\cr \noalign{\hrule}  
& Type && $npts$&& $L_1(\rho)$&& $L_1(p)$&& $L_1(v)$ && $r(\rho)$ &&
  $r(p)$ && $r(v)$ &\cr
  \noalign{\hrule}

&          && 100 && 3.72E-1 && 3.40E-1 && 4.25E-2 && 0.58 && 0.62
&& 0.66  &\cr
& Flux-splitting && 200 && 2.49E-1  && 2.20E-1 && 2.68E-2 && 0.61
&& 0.66  && 0.72   &\cr
& $O(\Delta x,\Delta t)$ && 400 && 1.63E-1 && 1.38E-1 && 1.62E-2 &&
0.65 && 0.698 && 0.75 \laststrut &\cr  
& (non-TVD)&& 800 && 1.03E-1 && 8.55E-2 && 9.64E-3 &&  0.70 && 0.73 && 0.85 &\cr
&          && 1600 &&  6.38E-02  && 5.14E-02 && 5.33E-03 && && &&
\laststrut &\cr 
  \noalign{\hrule}
&          && 100 && 0.37 && 0.33 && 4.23E-2  && 0.57 && 0.62 && 0.66 &\cr
& Godunov  && 200 && 0.24  && 0.22 && 2.67E-2 && 0.61 &&  0.66&& 0.72
&\cr
& $O(\Delta x,\Delta t)$ && 400 && 0.16 && 0.13 && 1.62E-2 && 0.65 &&
0.69 &&  0.75  \laststrut &\cr  
& (non-TVD)&& 800 && 0.10  && 8.51E-2 && 9.62E-3 && 0.70 && 0.73 && 0.85 &\cr 
&          && 1600 && 6.36E-2 && 5.12E-2 && 5.32E-3 && && &&\laststrut &\cr
  \noalign{\hrule}
&  w=-1  && 100 && 0.22 &&  0.23 && 1.89E-2  && 0.29 && 0.62 && 0.51  &\cr
& (Lax-Wend.)&& 200  && 0.18 && 0.15 && 1.32E-2 && 0.38 && 0.49 && 0.31  &\cr
& $O(\Delta x^2,\Delta t^2)$ && 400 &&0.14  && 0.11 && 1.06E-2 &&
2.09E-2 && 0.22 && -0.14 \laststrut &\cr  
& (non-TVD)&& 800 && 0.13 && 9.37E-2 && 1.17 E-2 && && && \laststrut &\cr
  \noalign{\hrule}
&          && 100 &&  0.28  && 0.25 && 2.73E-2 && 0.68 && 0.66  &&
0.75  &\cr
& $\beta =1$ && 200 && 0.17  && 0.15  && 1.61E-2  && 0.80 && 0.70 &&
0.83  &\cr
& $O(\Delta x^2,\Delta t^2)$ && 400 && 0.10 && 9.65E-2 && 9.08E-3  &&
0.78 && 0.71 && 0.79 
\laststrut &\cr  
& (TVD)    && 800 && 5.85E-2 && 5.90E-2 && 5.24E-3 &&  0.76 &&  0.73 && 0.86 &\cr
&          && 1600 && 3.43E-2  && 3.54E-2  && 2.88E-3 && && &&  \laststrut &\cr
  \noalign{\hrule}
&          && 100 &&  0.19 &&  0.14 && 1.95E-2  &&  0.89 && 0.91  &&  0.87 &\cr
& minmod   && 200 && 0.10 && 7.68E-2 &&  1.06E-2 && 0.87 && 0.94 &&
  0.89   &\cr 
& $O(\Delta x^2,\Delta t^2)$ && 400 && 5.82E-2  && 3.99E-2  && 5.72E-3
  &&  0.75 && 0.93 && 0.79 \laststrut &\cr  
& (TVD)    && 800 && 3.45E-2 && 2.09E-2 && 3.31E-3 && 0.85 && 0.97  && 0.99 &\cr 
&          && 1600 && 1.91E-2  && 1.06E-2  &&  1.67E-3  && && &&  \laststrut &\cr
\noalign{\hrule\vskip4pt}
  \noalign{\vskip5pt} }} 
\label{Comparing L1 errores for shock wave}
\end{table}

\newpage

\newpage

\newpage

\newpage

\newpage

\newpage

\newpage

\newpage

\end{document}